\documentclass[%
 reprint,
superscriptaddress,
 amsmath,amssymb,
 aps,
]{revtex4-2}
\usepackage{color}
\usepackage{gensymb}
\usepackage{graphicx}
\usepackage{dcolumn}
\usepackage{bm}
\usepackage{hyperref}
\usepackage{epstopdf}
\usepackage{float}
\usepackage{listings}
\usepackage{times,mathptm,bm}
\usepackage{xcolor}
\usepackage[export]{adjustbox}
\usepackage{upgreek}
\usepackage{chemformula} 

\usepackage{lineno}
\usepackage{siunitx}
\usepackage{array}

\catcode`\@=11 %

\font\uwavefont=lasyb10 scaled 652

\def\uwave{%
  \bgroup
    \markoverwith{%
      \lower3.5\p@\hbox{\uwavefont\char58}%
    }%
  \ULon
}

\begin{document}

\title{Drop-drop coalescence: a simple crossover function between inertial and viscous dynamics}

\author{Kaili Xie}
\email{k.xie2@uva.nl}
\affiliation{Van der Waals-Zeeman Institute, Institute of Physics, University of Amsterdam, 1098XH Amsterdam, The Netherlands.}

\author{Marie Corpart}
\affiliation{Van der Waals-Zeeman Institute, Institute of Physics, University of Amsterdam, 1098XH Amsterdam, The Netherlands.}

\author{Antoine Deblais}
\affiliation{Van der Waals-Zeeman Institute, Institute of Physics, University of Amsterdam, 1098XH Amsterdam, The Netherlands.} 

\author{Daniel~Bonn}
\affiliation{Van der Waals-Zeeman Institute, Institute of Physics, University of Amsterdam, 1098XH Amsterdam, The Netherlands.}
\date{\today}

\begin{abstract}
The coalescence of liquid drops is a fundamental process that remains incompletely understood, particularly in the intermediate regimes where capillary, viscous, and inertial forces are comparable. Here, we experimentally investigate the dynamics of drop-to-drop coalescence during the transition between viscous and inertial regimes using high-speed imaging. Our results reveal that the liquid bridge between droplets shows power-law growth with exponents between 1/2 and 1 during drop coalescence. We propose a novel scaling approach using a dimensionless crossover function that smoothly transitions between viscous and inertial limits. This simple approach, inspired by previous work on drop impact, successfully collapses the experimental data for a wide range of liquid viscosities and coalescence times onto a single master curve. We further compare our results with recent theoretical models and demonstrate how our approach complements and extends current understanding in the crossover of drop coalescence. This study contributes to both the fundamental physics of drop coalescence and its practical applications in various industrial processes.

\end{abstract}

\pacs{Valid PACS appear here}
\keywords{Hydrodynamics, Coalescence, Viscous-to-inertial transition, Droplets}
\maketitle

The coalescence of liquid drops is a fundamental process that plays a crucial role in many natural phenomena and industrial applications, from raindrop formation to ink-jet printing and emulsion stability \cite{eggers2008physics,derby2010inkjet,lohse2022fundamental,mcclements2004food,nannette2024thin}. Understanding the dynamics of liquid drop coalescence is essential for optimizing processes that involve liquid-liquid interactions and for developing advanced materials and technologies \cite{glasser2019tuning,seemann2012droplet}. When two liquid drops come into contact, a thin liquid bridge forms between them and rapidly expands, driven by surface tension forces. The liquid bridge grows from microscopic to macroscopic scales, after which the two drops merge into one.
This growth of the bridge has been shown to exhibit distinct regimes (viscous and inertial), depending on the relative importance of viscous and inertial forces \cite{paulsen2011viscous,eggers1999coalescence,aarts2005hydrodynamics, eggers2024coalescence}. The complexity of the crossover between these two regimes currently hampers a comprehensive understanding of the coalescence process as well as the development of practical applications thereof, such as droplet size control \cite{lohse2022fundamental,tran2013tubes}. In this Letter, we propose a simple scaling function to approximate the viscous-to-inertial transition in drop-drop coalescence, avoiding overcomplicated calculations. 

At very early times or for viscous fluids,
the drop-drop coalescence dynamics is dominated by a balance between viscous and capillary forces, resulting in a linear growth in time $t$ with a logarithmic correction \cite{eggers1999coalescence}, $R \sim t \ln t$.
In the later stages or for low-viscosity fluids, inertial forces dominate the bridge's growth, leading to a second self-similar regime characterized by $R \sim t^{1/2}$. 
While the asymptotic behaviors of both viscous and inertial regimes are well-established, the nature of the transition between them remains not fully understood, particularly in terms of its universality. The presence of logarithmic corrections to the viscous regime adds further complexity, obscuring how these corrections evolve during the viscous-to-inertial crossover, in which the capillary, inertial, and viscous forces all play significant roles.

From a practical standpoint, experimental observations of the crossover regimes in drop coalescence are challenging, especially for the low-viscosity fluids. Whether viscous or rather inertial forces dominate the coalescence dynamics is determined by the local Reynolds number, $Re$ = $\rho \gamma R^2 / (R_0 \eta^2)$, which depend on the fluid's physicochemical properties: surface tension ($\gamma$), dynamic viscosity ($\eta$), and density ($\rho$), as well as the initial drop radius ($R_0$) \cite{paulsen2011viscous}. 
For the coalescence of water drops ($\gamma \approx$ 70 mN/m, $\rho$ = 1 g/cm$^3$, $\eta$ = 1 mPa$\cdot$s, $R_0$ = 1 mm), the crossover occurs at a bridge
radius $R \approx \SI{3.7}{\micro m}$, corresponding to a time scale of $R/(\gamma / \eta) \approx$ 10$^{-8}$ s.
These extremely small length and time scales for water and other low-viscosity fluids make it very difficult to probe the viscous-to-inertial transition regime. Increasing the fluid's viscosity or decreasing its surface tension can significantly increase the crossover's length and time scales.

\begin{figure*}[htbp]
   \centering
   \includegraphics[width=2\columnwidth]{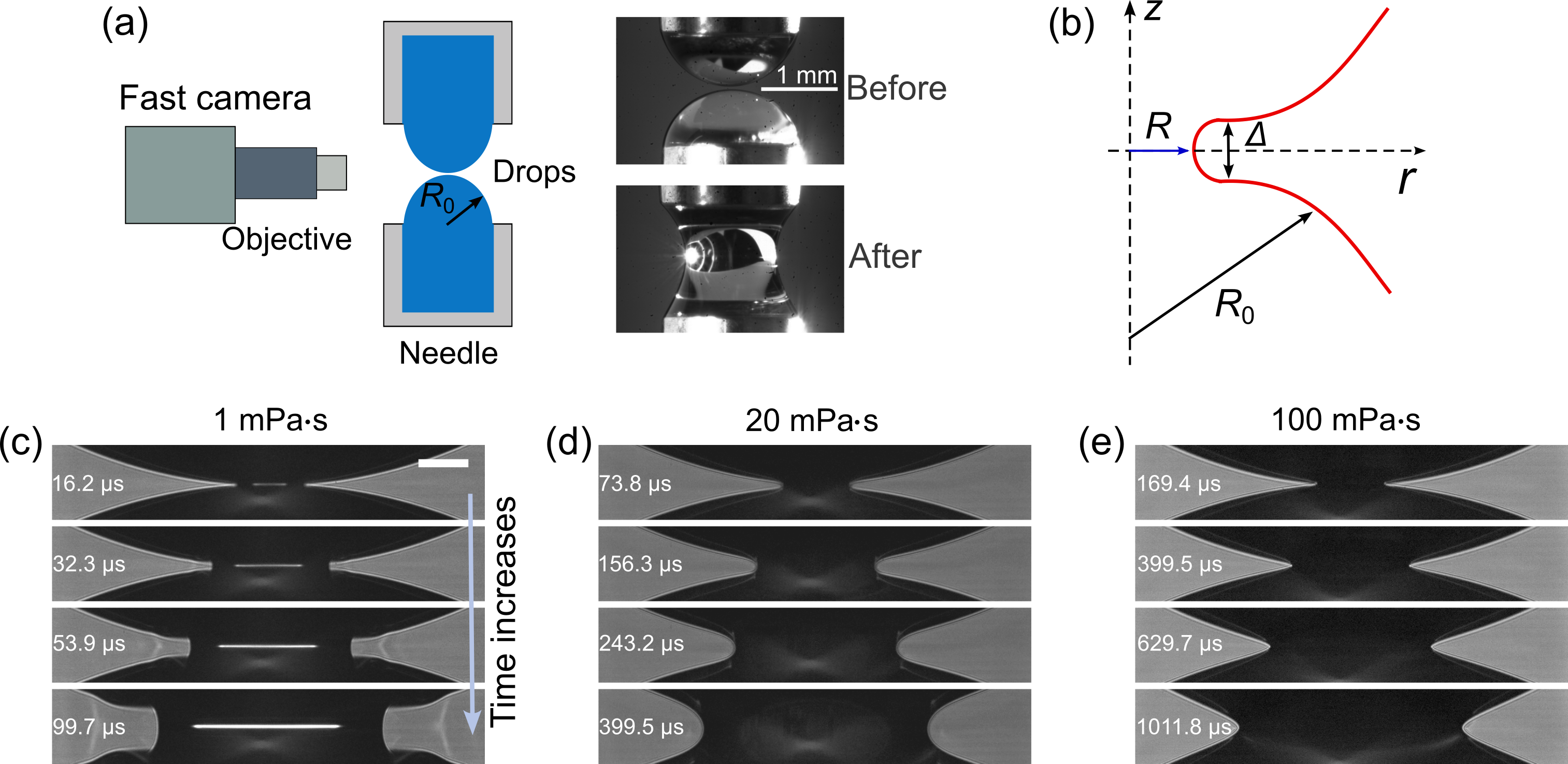}
    \caption{\textbf{Experimental setup.} (a) Schematic experimental setup to study drop-drop coalescence (left), using two vertically aligned needles carrying droplets and a fast imaging system. Typical images (right) show the moments before and after drop coalescence. (b) Schematic representation of the liquid bridge between two drops during coalescence. $R$ denotes the radius of the liquid bridge, $R_0$ the droplet's initial radius, and $\Delta$ the height of the bridge. (c)-(e) Close-up image sequences of coalescing drops with viscosities of 1 mPa$\cdot$s (water), 20 mPa$\cdot$s, and 100 mPa$\cdot$s, respectively. With increasing fluid viscosity, the coalescing drop shape changes from a \textit{V}-shape to \textit{U}-shape. Scale bar: $\SI{100} {\micro \meter}$}
    \label{fig:exp_setup}
\end{figure*}

Previously, Thoroddsen \textit{et al.} \cite{thoroddsen2005coalescence} studied the coalescence dynamics between pendent and sessile drops using various water-glycerol mixtures and alcohol drops, with fluid viscosities up to approximately 500 mPa$\cdot$s. Aarts \textit{et al.} \cite{aarts2005hydrodynamics} used the similar geometry and extended the viscosities of fluids up to 1000 mPa$\cdot$s. Both experimental studies show that the growth of the liquid bridge in the experiments agrees well with the inertial scaling in theory, $R \sim t^{1/2}$ while the bridge growth for viscous fluids follows a linear scaling $R \sim t$ without logarithmic correction. Nevertheless, none of these studies addressed a scaling that predicts bridge growth for fluids with intermediate viscosities, where the crossover occurs. 

Later on, Paulsen \textit{et al.} \cite{paulsen2011viscous,paulsen2013approach} proposed a fitting crossover function, $R/R_c = 2 \left( \frac{1}{t/t_c} + \frac{1}{\sqrt{t/t_c}} \right)^{-1}$ based on experimental observations, where the crossover length scale $R_c$ and timescale $t_c$ are the free parameters for each fluid to produce the best collapse of the experimental data. Experimentally, they found $R_c \sim R_0 Oh$ and $t_c \sim t_v Oh$, where $t_v = \eta R_0/\gamma$ is the viscous time, and $Oh = \eta / \sqrt{\rho \gamma R_0}$ is the Ohnesorge number, a nondimensional fluid viscosity. The crossover characteristics, $R_c$ and $t_c$ were further confirmed by Xia \textit{et al.} \cite{xia2019universality} using an exact analytical solution for the flow in the vicinity of the liquid bridge, giving a more comprehensive physical description. Although experimental data can be rescaled well using their crossover functions, this collapse should be considered an approximation, as a universal crossover function may not exist due to the logarithmic corrections to the viscous regime \cite{eggers2024coalescence}. It is crucial here to recognize that the concept of a `definitive' crossover function is fundamentally problematic. The presence of logarithmic terms in reality for the viscous regime means that the Ohnesorge number ($Oh$) cannot be strictly eliminated, and none of these crossover functions take the logarithmic terms into account.

In this Letter, we propose a simple crossover function as an alternative to describe the viscous-to-inertial transition in drop-drop coalescence, confirmed by our experiments covering a large range of fluid viscosities, and by other experimental results in the literature. Our crossover function smoothly interpolates between the two limiting behaviors (viscous and inertial) and serves as an approximation rather than an exact representation. This approach avoids overemphasizing too complicated calculations of the flow field near the liquid bridge to obtain a specific crossover function. This contrasts with the work of Xia \textit{et al.} \cite{xia2019universality}, who suggest a universal form. A key point of contention, however, is their assertion that even minimal inertia eliminates logarithmic corrections to the bridge growth, a claim that contradicts established understanding in the field. Our approach, therefore, aims to strike a balance between simplicity and accuracy, by employing a crossover function that is significantly simpler than the calculations presented by Xia \textit{et al.} \cite{xia2019universality} and the phenomenological model of Paulsen \textit{et al.} \cite{paulsen2011viscous}. Further, we aim to provide an alternative perspective on understanding coalescence dynamics, which may have implications for a wide range of applications, such as droplet size control \cite{mazutis2012selective, kooij2022self, bach2004coalescence, lohse2022fundamental, Chesters1991modelling}.

\textbf{\textit{Experiments -}} We study the drop-drop coalescence of deionized water and silicone oils, covering the viscosity range from 1 to 1000 mPa$\cdot$s. The experimental setup utilizes two vertically aligned metal needles with millimeter-scale tips. They are initially positioned several millimeters apart (Fig. \ref{fig:exp_setup}(a)). Each needle is connected to an independent syringe pump (New Era NE-1010). To mitigate undesired electrostatic effects due to surface charges on the drops, an electrostatic ion gun is used to discharge the drop surfaces before triggering coalescence. The upper drop is then brought down while the bottom drop remains stationary. Once they are close enough, the upper drop is enlarged slowly by pumping the fluid at a flow rate of $\SI{100} {\micro L/h}$, which corresponds to an approach velocity of approximately $\SI{20} {\micro \meter/s}$. In this approach we ignore inertial effects during the merging of the hemispherical drops formed at the needle tips.
Experiments reveal that the flow rate has no significant influence on the coalescence dynamics. After the coalescence is complete, a stable meniscus forms between these two drops (Fig. \ref{fig:exp_setup}(a)). A high-speed camera (Phantom TMX7510) equipped with a microscopic objective (20$\times$, WD = 20 mm, Edmund Optics) is used to capture the coalescence dynamics at 230,263 frames per second. The spatial resolution in imaging is $\SI{1.09} {\micro m /pix}$. The growth of the liquid bridge with time is then obtained using a home-made Matlab algorithm.
To determine the initial contact moment in coalescence $t_0$, we apply an extrapolation method by fitting the experimental data using a power law $R = \beta (t - t_0)^{\alpha}$, where $\alpha$ and $\beta$ are fitting parameters (see Appendix A). This method provides a precise estimate of $t_0$, which is difficult to determine directly from the recorded images due to optical opacity at the site where the liquid bridge initially forms \cite{aarts2005hydrodynamics, thoroddsen2005coalescence}. 
Therefore, in the following, we define time as $\tau = t - t_0$.

\textbf{\textit{Inertial, intermediate, and viscous regimes -}} 
The coalescence process, initiated when two droplets come into contact, is characterized by the formation and growth of a liquid bridge between the drops with a minimum radius $R$ and height scale $\Delta$ (Fig. \ref{fig:exp_setup}(b)). Depending on the relative dominance of inertial and viscous forces over capillary forces, the size and shape of the bridge evolve differently, as exemplified in Fig. \ref{fig:exp_setup}(c)-(e). For low viscosities (e.g., water, 1 mPa$\cdot$s, Fig. \ref{fig:exp_setup}(c)), the coalescence proceeds very rapidly, exhibiting a \textit{U}-shaped contour (half of the liquid bridge). In this case, the separation of the meniscus $\Delta \sim R^2/R_0$. By balancing the capillary force with the inertial force, $\gamma / \Delta \sim \rho u^2$ (where $u \sim R/ \tau$), one finds the nondimensional scaling for the bridge growth in the inertial regime, 
\begin{equation}
    \centering
    \frac{R}{R_0} = C_0 \left(\frac{\tau}{\tau_i}\right)^{1/2},
    \label{eq:inertial_scaling}
\end{equation}
where $\tau_i = \sqrt{\rho R_0^3 /\gamma }$ is the inertial time, and $C_0$ is a prefactor. 

For higher fluid viscosities (e.g., silicone oil, 100 mPa$\cdot$s), the liquid bridge grows more slowly and exhibits a \textit{V}-shaped contour (Fig. \ref{fig:exp_setup}(e)). In this case, the coalescence dynamics can take place over a wide range of scales (between $R^3/R_0$ and $R$) \cite{eggers1999coalescence,eggers2024coalescence}, leading to a more complex asymptotic solution that balances the capillary force with the viscous force. The nondimensionl scaling for this viscous regime is expressed, 
\begin{equation}
    \centering
    \frac{R}{R_0} = D_0 \frac{\tau}{\tau_v}, 
    \label{eq:viscous_scaling}
\end{equation}
with a prefactor $D_0$ and viscous time $\tau_v$. The full theoretical prediction for three-dimensional drop coalescence suggests $D_0 \approx -1/\pi \ln(\tau/\tau_v)$ with a leading-order logarithmic term \cite{eggers1999coalescence}.

\begin{figure}[htbp]
    \centering
   \includegraphics[width=0.7\columnwidth]{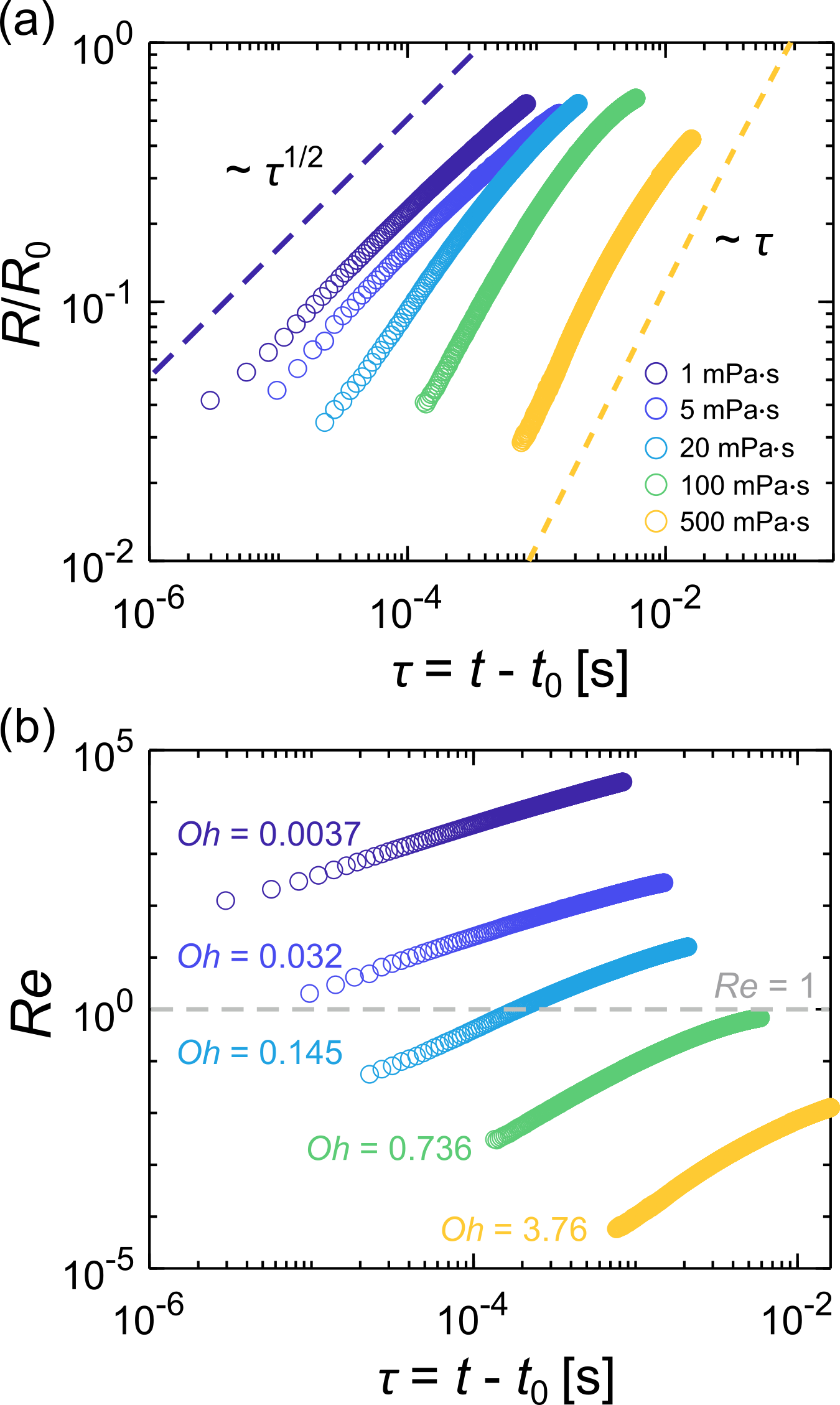}
    \caption{\textbf{Growth of the liquid bridge during coalescence.} Normalized bridge size (a) and local Reynolds number (b) as a function time $\tau = t - t_0$ for different fluid viscosities (or $Oh$ numbers). Here, $t_0$ is the moment of initial contact, which is determined using an extrapolation fitting with a power-law function (see Appendix A). In all experiments, the two coalescing drops have similar initial radius $R_0$. } 
    \label{fig:transition}
\end{figure}

Eqs. \ref{eq:inertial_scaling} and \ref{eq:viscous_scaling} have been previously confirmed by various experimental and simulation works \cite{aarts2005hydrodynamics,thoroddsen2005coalescence, xu2022bridge,deblais2024early,case2009coalescence,wu2004scaling, paulsen2011viscous,sprittles2012coalescence,anthony2020initial,anthony2023sharp}. However, none of these scalings can describe the intermediate regime where the capillary, inertial, and viscous forces are all comparable during coalescence (Fig. \ref{fig:transition}(a)). 

We study the effects of viscous and inertial forces on these power laws by systematically varying the fluid viscosities from 1 mPa$\cdot$s ($Oh$ = 0.0037) to 500 mPa$\cdot$s ($Oh$ = 3.7). 
Fig. \ref{fig:transition}(a) shows the evolution of the liquid bridge as a function of fluid's viscosity, revealing a crossover between viscous ($R/R_0 \sim \tau$) and inertial ($R/R_0 \sim \tau^{1/2}$) coalescence regimes.

\begin{figure*}[htbp]
    \centering
   \includegraphics[width=1.5\columnwidth]{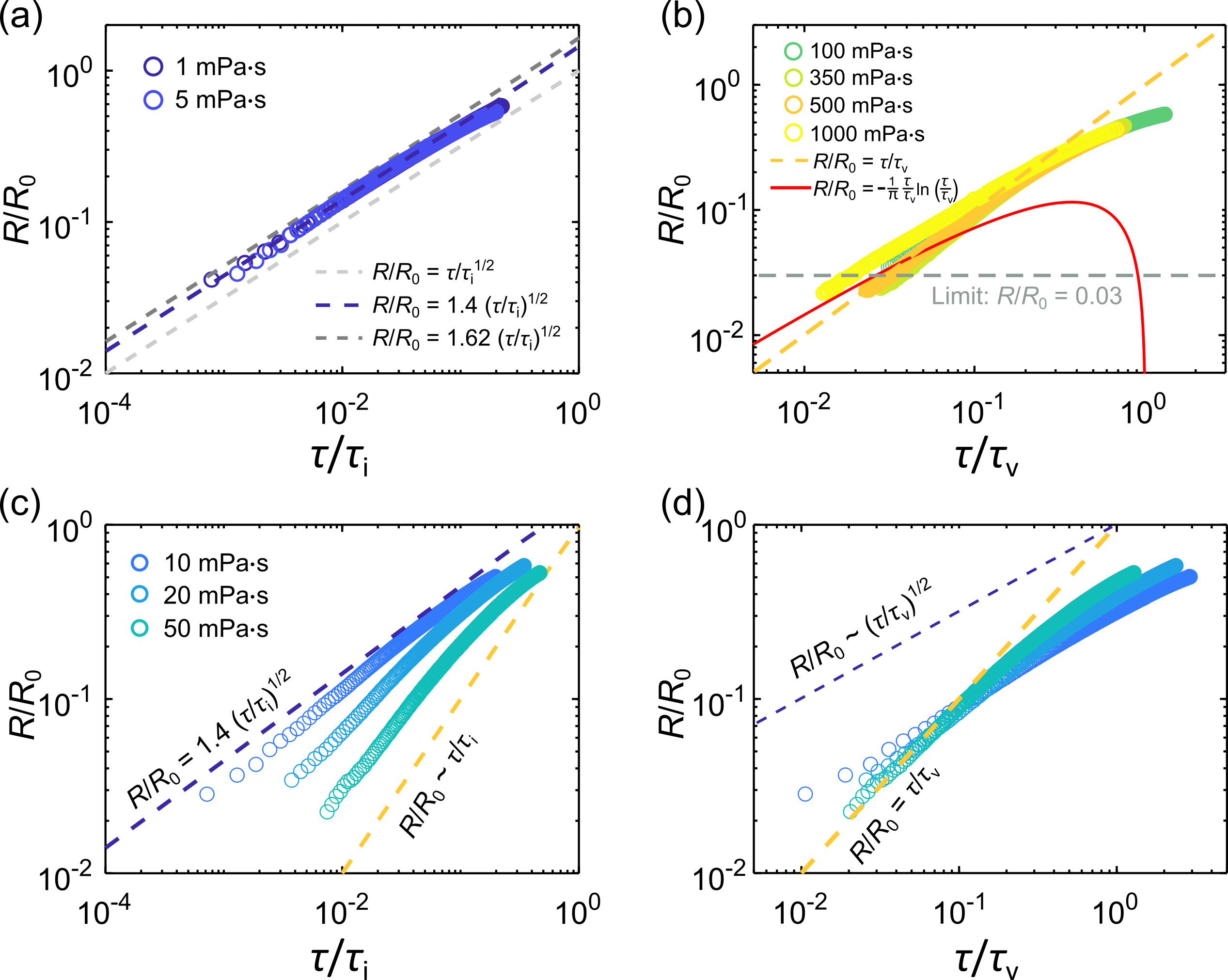}
    \caption{\textbf{Evolution of the liquid bridge radius during drop coalescence in different regimes.} (a) Inertial regime. The bridge size is described well by the inertial scaling law (Eq. \ref{eq:inertial_scaling}). Different prefactors are shown for comparision. (b) Viscous regime. The bridge size follows well the linear prediction (Eq. \ref{eq:viscous_scaling}) but deviates from the logarithmic scaling law. The horizontal dashed line indicates the limit of validity for the theory with logarithmic correction. (c,d) Intermediate regime. The bridge size grows with power laws between $\sim t$ and $\sim t^{1/2}$ for the intermediate viscosities. } 
    \label{fig:scaling_compare}
\end{figure*}

To further elucidate the distinct coalescence regimes, we compare our experimental results with the scaling laws in Eqs. \ref{eq:inertial_scaling} and \ref{eq:viscous_scaling}, as shown in Fig. \ref{fig:scaling_compare}.
In the inertial regime, typically observed for low-viscosity fluids or in the very late stages of coalescence, the bridge size follows the inertial scaling law (Eq. \ref{eq:inertial_scaling}), yielding a prefactor $C_0$ = 1.4 $\pm$ 0.2. This value agrees well with the prefactors obtained by others \cite{aarts2005hydrodynamics,wu2004scaling, beaty2024inertial}, although it is lower than the value of 1.62 predicted numerically by Duchemin \textit{et al.} \cite{duchemin2003inviscid} (Fig. \ref{fig:scaling_compare}(a)). In contrast, in the viscous regime, prevalent in high-viscosity fluids or early stages of coalescence, our experimental data suggests linear growth with a prefactor $D_0 \approx$ 1 (Eq. \ref{eq:viscous_scaling}) rather than linear growth with a logarithm correction (see Fig. \ref{fig:scaling_compare}(b)). The discrepancy arises because the validity of the asymptotic theory with logarithmic correction \cite{eggers1999coalescence} applies only to small bridge sizes, $R/R_0 <$ 0.03, which is close to the bottom limit of the bridge radius in our experiments. The logarithmic correction may be present but overshadowed by azimuthal effects when $R/R_0 >$ 0.03, and is therefore not observed in our or previous drop-drop coalescence experiments \cite{aarts2005hydrodynamics,thoroddsen2005coalescence}. However, it has been borne out by recent numerical simulations \cite{sprittles2014parametric,anthony2020initial}. Simulation results \cite{anthony2020initial} show that the bridge growth of a Stokes fluid ($Oh \rightarrow \infty$) follows the theoretical prediction with the logarithmic correction \cite{eggers1999coalescence} in the very early stage (up to $\tau_v \approx$ 10$^{-2}$ and $R/R_0 \approx$ 0.02), but deviates from linear scaling when the logarithmic correction is omitted. However, in the later stage ($\tau_v >$ 10$^{-2}$), there is no significant difference between the Stokes simulation and the linear scaling, which is the time range we probed in our experiments (Fig. \ref{fig:scaling_compare}(b)).

Of particular interest is the intermediate regime we observe power-law growth of the bridge size with exponents between 1/2 and 1, bridging the gap between the inertial and viscous limits (Fig. \ref{fig:scaling_compare}(c),(d)). The experimental results show that the intermediate regime starts at  $\eta \leq$ 50 mPa$\cdot$s ($Oh \leq 0.36$) and ends at $\eta \geq$ 10 mPa$\cdot$s ($Oh \geq 0.07$). Clearly, for the fluid with a viscosity of 20 mPa$\cdot$s ($Oh = 0.14$), the bridge growth follows neither the inertial nor the viscous scaling laws. In this case, $Re$ crosses unity throughout the entire coalescence process (Fig. \ref{fig:transition}(b)), further confirming the transition regime.

\medskip
\textbf{\textit{Crossover dynamics -}} 
We now introduce a simple scaling approach to describe the bridge growth in coalescence across the different regimes. In our analysis, we employ the crossover rescaling method, which was first proposed by Eggers \textit{et al.} \cite{eggers2010drop} for drop impact. Later, this approach was extended to describe the crossover between inertial-capillary and inertial-viscous drop impact regimes for both Newtonian and non-Newtonian fluids \cite{laan2014maximum, jiang2024frozen}. We follow this approach \cite{laan2014maximum} by introducing a function $f(P)$ that serves to distinguish between the viscous and inertial regimes. Therefore, the nondimensional bridge growth should follow $R/R_0 = f(P) \tau /\tau_v$ for both regimes. The smooth crossover between these two regimes is then constructed by using a Padé approximant \cite{press2007numerical}, $f(P) = \sqrt{P}/(A + \sqrt{P})$, where $A$ is a constant. Then, the crossover scaling in drop coalescence can be expressed as,
\begin{equation}
    \centering
    \frac{R}{R_0} = \frac{\sqrt{P}}{A + \sqrt{P}} \left( \frac{\tau}{\tau_v}\right),
    \label{eq:crossover_scaling}
\end{equation}
with only one unknown dimensionless parameter $P$. If $A \ll \sqrt{P}$, then $f(P) \approx$ 1, and Eq. \ref{eq:crossover_scaling} simplifies to be $R/R_0 \approx \tau/\tau_v$, recovering the viscous scaling of Eq. \ref{eq:viscous_scaling} with a prefactor $D_0$ = 1. If $A \gg \sqrt{P}$, Eq. \ref{eq:crossover_scaling} becomes $R/R_0 \approx \sqrt{P}/A (\tau/\tau_v)$, which should recover the inertial regime. Using an analogy to the inertial scaling in Eq. \ref{eq:inertial_scaling}, we find the parameter $\sqrt{P}$ must satisfy $\sqrt{P} = \tau_v /\sqrt{\tau_i \tau}$. This yields a prefactor $1/A$ in Eq. \ref{eq:crossover_scaling} in this limit. To be consistent, it requires $C_0 = 1/A$. From our experiments for inertial fluids, we can know $A \approx$ 0.71, and we set this constant for all fluids.
We check our scaling argument by comparing the parameter $\sqrt{P}$ with $A$ over time during coalescence in both viscous and inertial regimes (see Appendix B). Fig. \ref{figS:A_sqrtP} further confirms that our two-limit assumption, as mentioned above, is plausible and that the choice of the constant $A$ is reasonable.

\begin{figure}[ht]
    \centering
   \includegraphics[width=1\columnwidth]{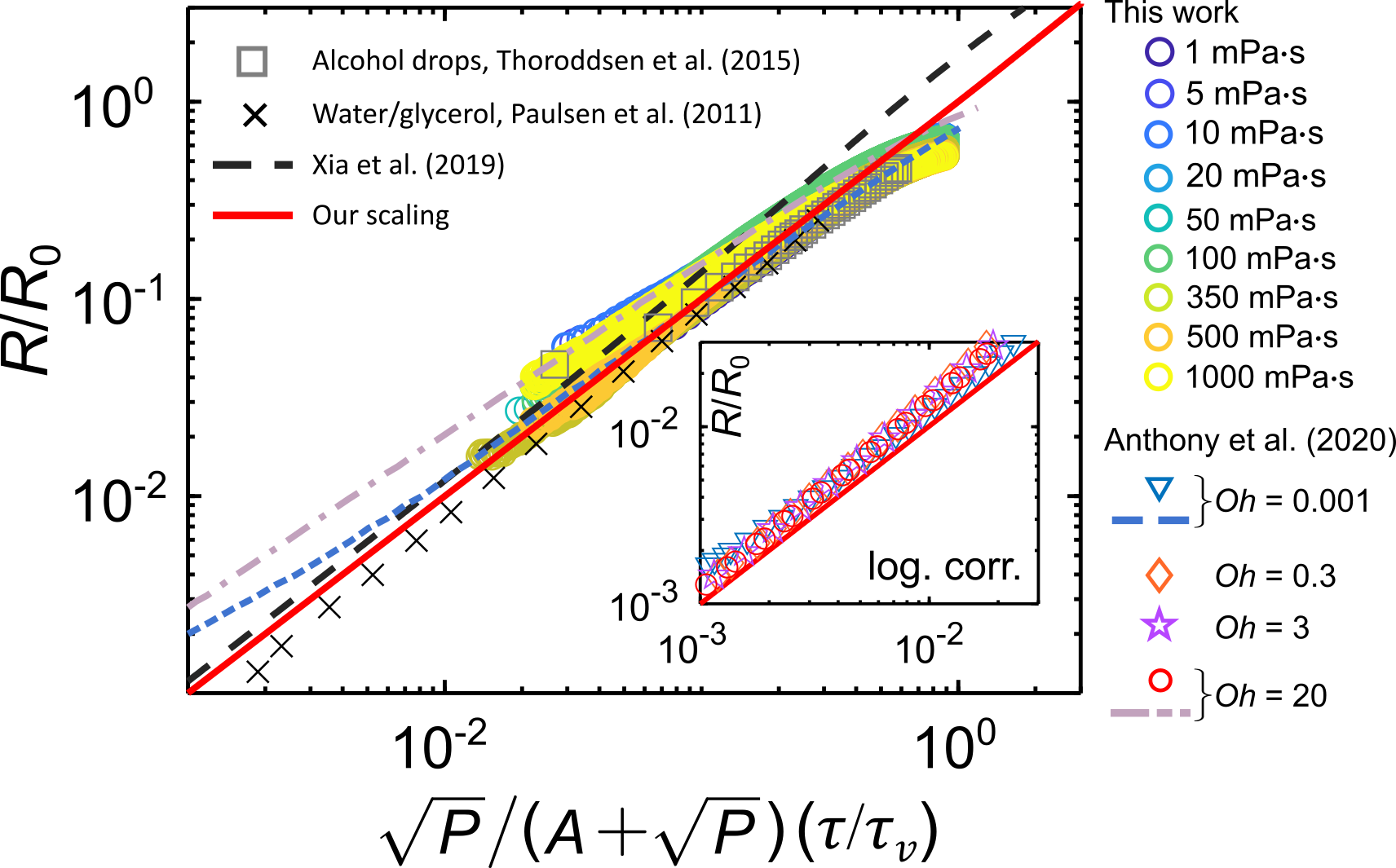}
    \caption{\textbf{Crossover scaling of drop-drop coalescence.} Liquid bridge size as a function of $\sqrt{P}/(A + \sqrt{P})(\tau/\tau_v)$ using our experimental data as well as data from \cite{thoroddsen2005coalescence,paulsen2011viscous} and theoretical predictions from \cite{xia2019universality}. Simulations by Anthony \textit{et al.} \cite{anthony2020initial} exhibit a better alignment with our crossover function with logarithm correction (inset) in the early stage of coalescence, where our drop-drop experimental observations are not applicable.} 
    \label{fig:crossover}
\end{figure}

We further compare our crossover scaling to experimental data in Fig. \ref{fig:crossover}. All our experimental data for liquids with viscosities ranging from 1 mPa$\cdot$s to 1000 mPa$\cdot$s, and experimental data from literature correspond very well with our crossover scaling law. In Fig. \ref{fig:crossover}, we also show that our simple scaling largely captures the recent crossover theory proposed by Xia \textit{et al.} \cite{xia2019universality}. 
The deviation between our crossover scaling and Xia's theory becomes pronounced as the liquid bridge approaches the initial drop size (Fig. \ref{fig:crossover}), where our scaling aligns more closely with the experimental data. 

The crossover scaling in Eq. \ref{eq:crossover_scaling} is derived asymptotically based on the linear growth of the bridge with a prefactor of $D_0 \approx$ 1. This scaling works very well when the bridge radius exceeds 0.03$R/R_0$ for all viscosities. However, it does not describe the crossover behavior in the early stages when logarithmic corrections are present. 
Although our experimental work cannot clearly capture these early-stage behaviors,
careful simulations enable us to verify our scaling.
Simulation work by Anthony \textit{et al.} \cite{anthony2020initial} shows a good agreement between our scaling and a fluid with a small Ohnesorge number ($Oh$ = 0.001) but a discrepancy for a $Oh$ = 20 in the range $R/R_0 <$ 0.03 (Fig. \ref{fig:crossover}), where the logarithmic corrections are believed to exist \cite{eggers1999coalescence}. To check this, we modify our scaling by incorporating a logarithmic term into the crossover parameter, $P$ = (-$\frac{\tau}{\pi\tau_v} \ln{\frac{\tau}{\tau_v}}$ )$^{-2} (\frac{\tau}{\tau_i})$. This log-corrected scaling accurately describes the simulation results for $R/R_0 <$ 0.03 across a range of $Oh$ = $0.001 < Oh < 20$ (see inset in Fig. \ref{fig:crossover}).

\textbf{\textit{Conclusion -}}
In summary, we have showed that a Padé approximation can be used to obtain a crossover function that successfully captures the essential features of the viscous-to-inertial transition in drop-drop coalescence.
We compared our simple crossover function with our own experiments across a range of liquid viscosities as well as with experimental and numerical results from the literature, revealing excellent correspondence. Moreover, our crossover explicitly incorporates the dependence on the $Oh$ number, while eliminating the need to determine the characteristic time and length scales at crossover. It also addresses and enables the description of dynamic behaviors with logarithmic corrections in the early stage of coalescence, which are not captured by current theoretical predictions. 

Quantifying the dynamics of the crossover between viscous and inertial regimes in drop coalescence is not only of fundamental scientific interest but also has practical implications for a wide range of applications. As such, our approach potentially allows for a more direct and simpler comparison of coalescence behaviors across different liquid systems, such as emulsion and non-Newtonian fluids \cite{bremond2008decompressing,tcholakova2006coalescence,dekker2022elasticity,chen2022probing}. This work may help shed light on such applications, including technologies to precisely control droplet size in microfluidics and aerosol generation.

\textbf{\textit{Acknowledgments -}}
This work was supported by Dutch Research Council NWO IPP Grant (Innovative Nanotech Sprays, ENPPS.IPP.019.001). K.X. gratefully acknowledges the funding support from Marie Skłodowska-Curie Actions Postdoctoral Fellowship (No.101150851).


\section*{APPENDIX A: Determination of the initial contact between drops}
\begin{figure}[ht]
\centering 
   \includegraphics[width=1\columnwidth]{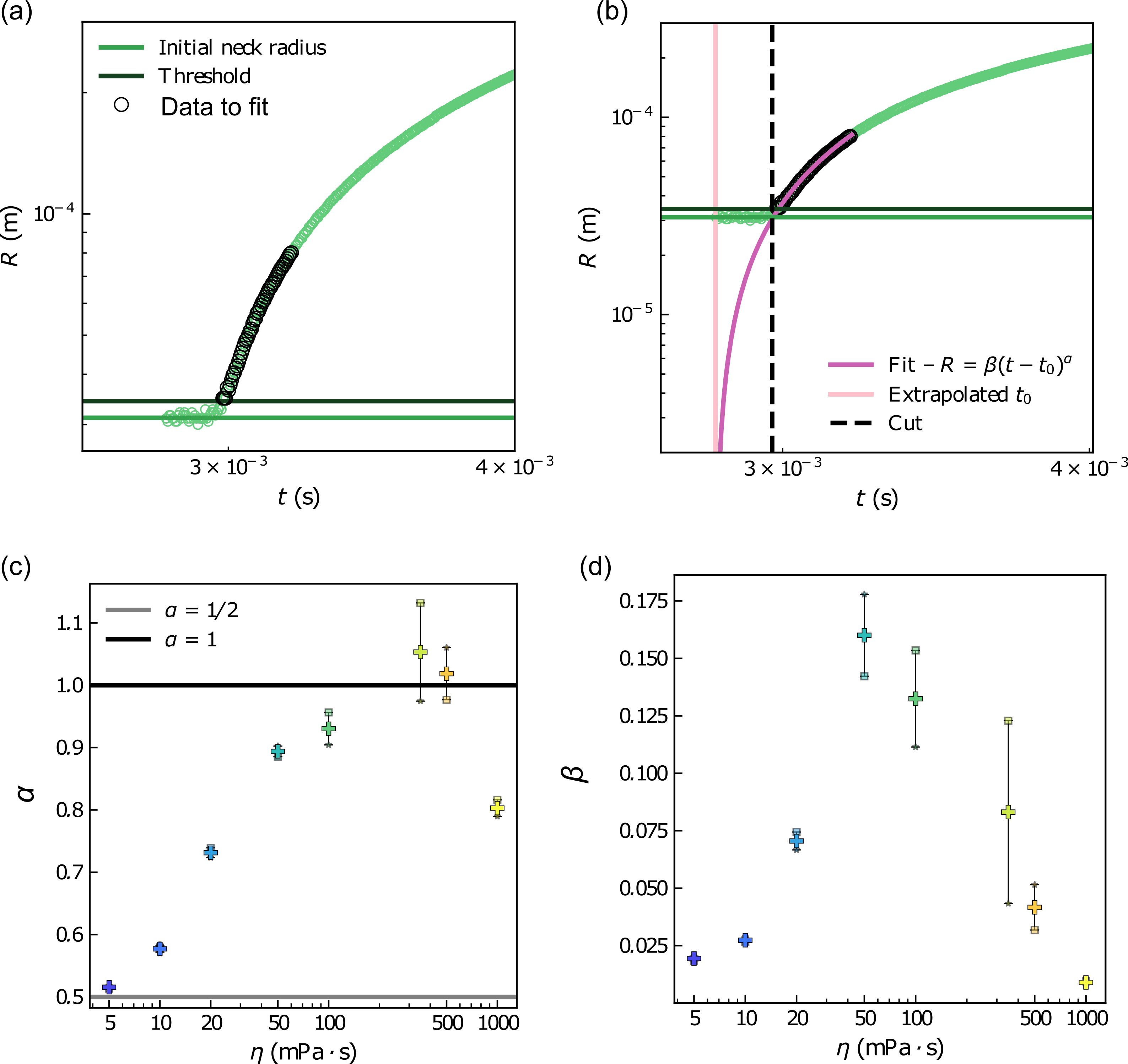}
  \caption{Determination of initial contact in drop coalescence. (a) A threshold applied to the raw experimental data to determine the data range used for fitting. (b) Fitting and extrapolation of experimental data. (c),(d) The exponents and prefactors from the power-law fitting for fluids with different viscosities.  } 
   \label{figS:t0_determine}
\end{figure}

To determine the initial contact during drop coalescence, we applied an extrapolation algorithm to fit the raw experimental data using a power-law function, $R = \beta (t - t_0)^{\alpha}$, where $R$ is the liquid bridge radius, $t - t_0$ is the time, and $\alpha$ and $\beta$ are fitting parameters. As shown in Fig. \ref{figS:t0_determine}(a),(b), a threshold was applied to determine the range of the data used for fitting, followed by fitting the data with the power-law function and extrapolation. The intersection between the fitted curve and the threshold is considered as the initial contact in drop coalescence. Therefore, the plateau in the early stage in our experimental data is cut off in all analyses. The exponents from the fitting for different fluids are between 0.5 and 1, covering the transition of viscous to inertial regimes (Fig. \ref{figS:t0_determine}(c)).

\section*{APPENDIX B: Comparison of the constant $A$ and the crossover parameter $\sqrt{P}$}
\begin{figure}[h!]
\centering 
   \includegraphics[width=0.8\columnwidth]{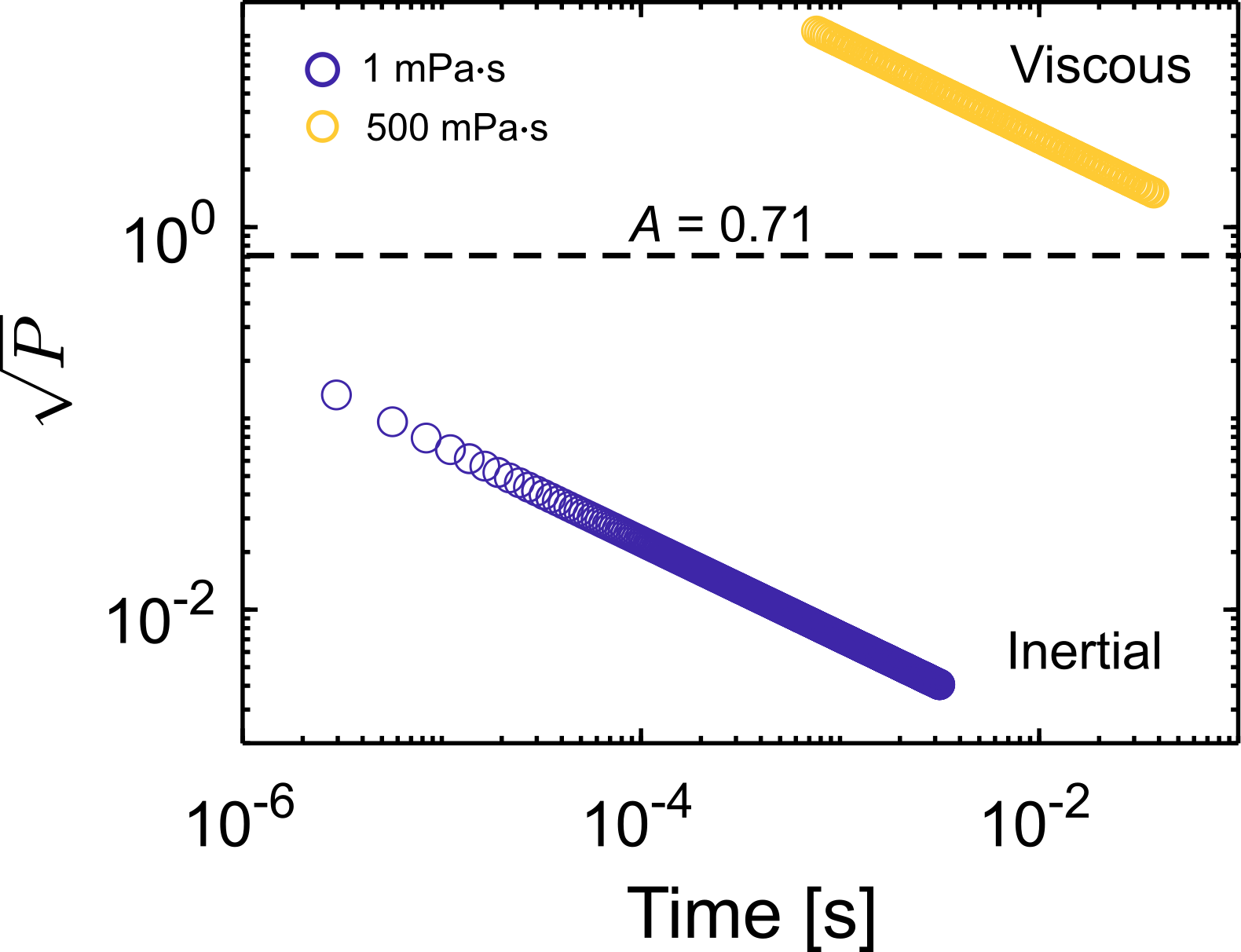}
  \caption{Comparison of the constant $A$ and crossover parameter $\sqrt{P}$ in two limits: viscous and inertial. In our crossover function, we set $A$ = 0.71. 
In the inertial regime, it shows $A \gg \sqrt{P}$, which satisfies our assumption in the scaling analysis. Similarly, for the viscous regime, $A \ll \sqrt{P}$, also meets the requirement.} 
   \label{figS:A_sqrtP}
\end{figure}

\newpage

\bibliographystyle{apsrev4-2} 
\bibliography{refs.bib}



\end{document}